\documentclass[a4paper]{article}

\usepackage{INTERSPEECH2022}
\usepackage{hyperref}
\usepackage{multirow}
\usepackage{svg}

\usepackage[absolute,overlay]{textpos}
\usepackage{xcolor}

\title{HiFi-VC: High Quality ASR-Based Voice Conversion}
\name{Anton Kashkin$^1$, Ivan Karpukhin$^1$, Svyatoslav Shishkin$^1$}
\address{
  $^1$Tinkoff}
\email{a.kashkin@tinkoff.ru, i.a.karpukhin@tinkoff.ru, s.shishkin@tinkoff.ru}

\begin{document}

\maketitle

\begin{textblock*}{10cm}(13cm,2.5cm)
{\color{gray}
   \large Submitted to INTERSPEECH 2022}
\end{textblock*}

\begin{abstract}
  The goal of voice conversion (VC) is to convert input voice to match the target speaker's voice while keeping text and prosody intact. VC is usually used in entertainment and speaking-aid systems, as well as applied for speech data generation and augmentation. The development of any-to-any VC systems, which are capable of generating voices unseen during model training, is of particular interest to both researchers and the industry. Despite recent progress, any-to-any conversion quality is still inferior to natural speech.
  
  In this work, we propose a new any-to-any voice conversion pipeline. Our approach uses automated speech recognition (ASR) features, pitch tracking, and a state-of-the-art waveform prediction model. According to multiple subjective and objective evaluations, our method outperforms modern baselines in terms of voice quality, similarity and consistency.
  
\end{abstract}

\section{Introduction}

Speech synthesis is aimed at predicting waveforms containing voices possessing desired properties \cite{ning2019ttsreview, sisman2020overview}. Text-to-speech and voice conversion are two main approaches to speech synthesis. In text-to-speech (TTS), an algorithm predicts a voice waveform based on the provided text. Sometimes extra information is available, such as emotion or tempo. Popular applications of TTS include voice assistants, smart home devices and natural interfaces. In voice conversion (VC), an algorithm converts the voice of one speaker to the voice of another speaker while preserving textual information and prosody. The most simple method of implementing VC is by combining automatic speech recognition (ASR) and TTS. However, the textual bottleneck of this approach can lead to losing important information about prosody. To tackle this problem, special VC algorithms were proposed, which are used in entertainment \cite{luo2020singing}, speaking-aid systems \cite{nakamura2012speakingaid}, data augmentation \cite{shahnawazuddin2020vcaugmentation} and anonymization \cite{maouche2020anonymization}.

In the last decade, significant progress was made in speech synthesis methods, including both TTS \cite{ning2019ttsreview} and VC \cite{sisman2020overview}. Most recent works perform voice conversion in three steps \cite{qian2019autovc,wang2021vqmivc}. First, they extract content and speaker features from input samples, then they generate a spectrogram, and finally convert it into a waveform using a vocoder. The core source of recent advances lies in using deep learning (DL) and large datasets. Several works make use of neural waveform encoders \cite{peddinti2015tdnn, ravanelli2018speaker, schneider2019wav2vec} and decoders \cite{oord2016wavenet}. Some methods apply ASR features along with voice pitch features for input coding \cite{polyak2019ttsskins,mottini2021voicy}. Many approaches exploit ideas from generative-adversarial networks (GAN) \cite{nguyen2021nvcnet,li2021starganv2}. Neural vocoders such as WaveGlow \cite{prenger2019waveglow} or HiFi GAN \cite{kong2020hifi} increased output speech quality compared to traditional approaches \cite{kawahara1999straight,perraudin2013fastgriffin}.

Voice conversion methods can be characterized by their level of complexity. Basic approaches convert the voice of one or multiple predefined speakers to the voice of a single target speaker \cite{sun2016manyone}. More flexibility can be achieved by using the so-called {\it many-to-many} methods, which are capable of converting voices in a closed set of speakers \cite{liu2021vqbn,polyak2019ttsskins}. However, while these methods excel at predicting a predefined set of voices, they are incapable of converting voices unseen during training, or can produce low quality speech. Finally, the most general approach to VC consists of converting arbitrary voices either seen or unseen during training \cite{nguyen2021nvcnet,wang2021vqmivc,kim2021assem}. This approach is usually called {\it any-to-any} or {\it zero-shot} voice conversion.


In this work, we propose a new high-quality any-to-any voice conversion system. Contributions of this paper can be summarized as follows:
\begin{enumerate}
    \item We propose a new conditional GAN architecture, which is capable of directly predicting a waveform from intermediate features. In particular, we adapt HiFi GAN \cite{kong2020hifi} vocoder for a general decoding task.
    \item We combine ideas from ASR-based content encoding with a GAN generation approach to achieve high-quality any-to-any voice conversion.
    \item We compare the proposed method with modern baselines using subjective and objective evaluations. According to our experiments, the proposed method achieves higher voice quality, similarity and consistency.
\end{enumerate}

\begin{figure}[t]
  \centering
  \includegraphics[width=0.95\linewidth]{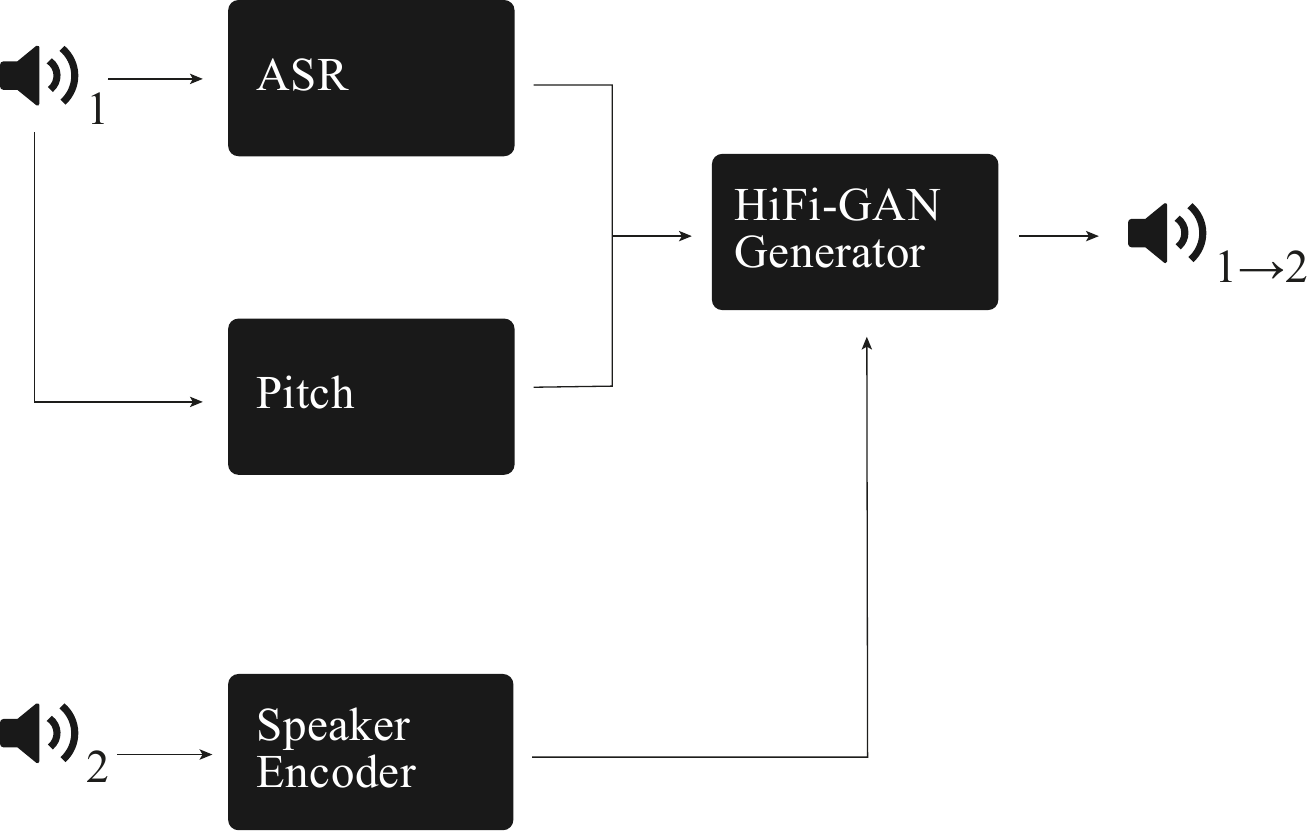}
  \caption{HiFi-VC inference pipeline. We use pretrained ASR bottleneck features and pitch tracker for encoding. Decoder and vocoder are joined into a single HiFi GAN-like model with an additional condition on speaker embedding.}
  \label{fig:inference}
\end{figure}

\begin{figure*}[t]
  \centering
  \includegraphics[width=\linewidth]{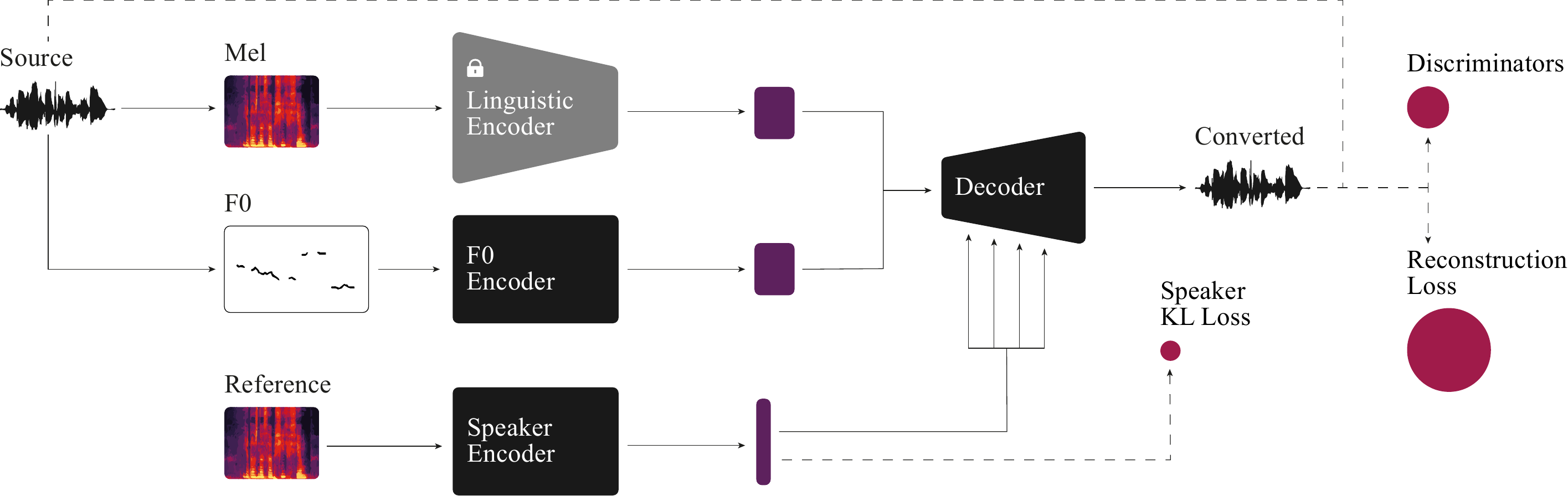}
  \caption{HiFi-VC training pipeline. Output waveform combines linguistic information and prosody from the source sample with reference timbre. The ASR model is used as a linguistic encoder, while the pitch encoder provides prosody information. Pretrained ASR model used in the content encoder is freezed during training.}
  \label{fig:pipeline}
\end{figure*}

\section{Related Works}

While VC is intrinsically close to both ASR and TTS, there are some problems unique to VC. To change voice features, it is necessary to extract speaker-agnostic content information from a source sample. Some methods are based on auto-encoder or variational auto-encoder architectures and make use of the bottleneck layer to eliminate speaker information from the latent space \cite{qian2019autovc,chou2019adain}. Additional normalization layers can be applied to further improve speaker and content features disentanglement \cite{chou2019adain}. Some works explicitly force disentanglement by either minimizing mutual information \cite{wang2021vqmivc}, by matching representations of the original and converted speech samples \cite{nguyen2021nvcnet} or by using discriminators \cite{sakamoto2021stargan}.

Many works make use of ASR model \cite{polyak2019ttsskins,mottini2021voicy,chen2021tvqvc}. Some methods apply ASR predictions (such as phonemes probabilities) for linguistic information coding \cite{mottini2021voicy,chen2021tvqvc}, while others utilize bottleneck features \cite{polyak2019ttsskins,liu2021vqbn}. When it comes to the latter, it was empirically observed that bottleneck ASR features contain little information about the source speaker. Another use of ASR is to design a training objective which minimizes the loss of linguistic information during conversion \cite{li2021starganv2}. One drawback of ASR-based coding is ASR's practical inability to encode prosody features. To handle this problem, many methods directly extract fundamental frequency (F0) from the source sample \cite{polyak2019ttsskins,kim2021assem,li2021starganv2}. Speaker information is usually eliminated from F0 by normalization \cite{kim2021assem,wang2021vqmivc}. Our method uses ASR and F0 encoder similar to TTS Skins \cite{polyak2019ttsskins}. F0 is additionally preprocessed by a trainable network similar to that in BNE-Seq2seqMoL \cite{liu2021f0prediction}.

Given content and speaker features, most methods decompose prediction into two steps: decoder and vocoder \cite{qian2019autovc, wang2021vqmivc,kim2021assem}. The goal of the decoder is to predict the Mel spectrogram of the output signal. The vocoder, on the other hand, converts the predicted spectrogram into an output waveform. The decoder can be implemented via RNN \cite{qian2019autovc}, Transformer \cite{chen2021tvqvc} or fully convolutional architecture \cite{sakamoto2021stargan}. Popular vocoders include STRAIGHT \cite{kawahara1999straight}, Griffin-Lim \cite{perraudin2013fastgriffin} and neural network models such as WaveGlow \cite{prenger2019waveglow} and HiFi GAN \cite{kong2020hifi}. To achieve better quality with neural approaches, an extra vocoder fine-tuning step is required after decoder training. 

Our method uses HiFi GAN decoder \cite{kong2020hifi}. Unlike previous works based on ASR and GAN \cite{liu2021vqbn}, we avoid the intermediate Mel spectrogram prediction step and directly produce a waveform from encoded features. Our method is also different in how speaker embedding affects conversions. In most previous works, speaker embedding is used in the decoder for spectrogram prediction \cite{qian2019autovc,liu2021vqbn,wang2021vqmivc}. In contrast, we directly condition GAN residual blocks with speaker embedding during generation. Our approach excludes a separate vocoder and simplifies conversion pipeline. Therefore, our method does not need a vocoder fine-tuning step after decoder training. Some previous works, such as VCRSS \cite{liu2021vqbn} and NVC-Net \cite{nguyen2021nvcnet}, also avoid using separate vocoder. Unlike VCRSS, we use a GAN decoder for better waveform prediction. Our approach is also defferent from NVC-Net, as we use the ASR encoder and a different GAN architecture.

\begin{table*}[ht]
  \caption{Many-to-many voice quality and similarity mean opinion scores (MOS) for the proposed method and baselines. "F" and "M" correspond to different gender combinations of source and reference voices. Standard deviation (STD) of the voice quality in all studies is less than $0.19$. STD of similarity is less than $0.26$. Ground truth score is obtained using original records from the dataset.}
  \label{tab:mos}
  \centering
  \begin{tabular}{l|ccccc|ccccc}
    \toprule
\multirow{2}{*}{\textbf{Model}} & \multicolumn{5}{|c|}{\textbf{Many-to-Many Voice Quality $\uparrow$}} & \multicolumn{5}{c}{\textbf{Many-to-Many Similarity $\uparrow$}} \\
& \textbf{F2F} & \textbf{F2M} & \textbf{M2M} & \textbf{M2F} & \textbf{Mean} & \textbf{F2F} & \textbf{F2M} & \textbf{M2M} & \textbf{M2F} & \textbf{Mean} \\
\midrule
Ground Truth & 4.30 & N/A & 4.35 & N/A & 4.33 & 4.37 & N/A & 4.44 & N/A & 4.40 \\
\midrule
AutoVC\cite{qian2019autovc} & 2.22 & 2.14 & 2.27 & 2.15 & 2.20 & 1.59 & 1.66 & 1.73 & 1.47 & 1.61 \\
VQMIVC\cite{wang2021vqmivc} & 3.93 & 3.69 & 3.74 & 3.78 & 3.78 & 2.97 & 3.10 & 3.19 & 2.97 & 3.06 \\
NVC-Net\cite{nguyen2021nvcnet} & 3.73 & 3.17 & 3.71 & 3.35 & 3.49 & 3.91 & 3.79 & 3.83 & 3.71 & 3.81 \\
\midrule
HiFi-VC (Proposed) & \bf 4.10 & \bf 4.09 & \bf 4.11 & \bf 4.01 & \bf 4.08 & \bf 4.03 & \bf 4.17 & \bf 4.11 & \bf 3.99 & \bf 4.08 \\

\bottomrule
\end{tabular}
\end{table*}

\begin{table}[ht]
  \caption{Many-to-many word error rate (WER), character error rate (CER) and prosody quality in terms of Pearson correlation coefficient (PCC) between source and predicted pitch features.}
  \label{tab:objective}
  \centering
  \begin{tabular}{l|ccc}
    \toprule
\multirow{2}{*}{\textbf{Model}} & \multicolumn{3}{|c}{\textbf{Many-to-Many}} \\
& \textbf{WER (\%) $\downarrow$} & \textbf{CER (\%) $\downarrow$} & \textbf{PCC $\uparrow$} \\
\midrule
AutoVC\cite{qian2019autovc} & 85.1 & 58.1 & 0.22 \\
VQMIVC\cite{wang2021vqmivc} & 32.5 & 16.9 & 0.51 \\
NVC-Net\cite{nguyen2021nvcnet} & 37.9 & 21.4 & 0.42 \\
\midrule
HiFi-VC (Proposed) & \bf 14.6 & \bf 6.4 & \bf 0.60 \\

\bottomrule
\end{tabular}
\end{table}

\begin{table*}[ht]
  \caption{Any-to-any voice quality and similarity mean opinion scores (MOS) for the proposed method and baselines. "F" and "M" correspond to different gender combinations of source and reference voices. Standard deviation (STD) of the voice quality in all studies is less than $0.23$. STD of similarity is less than $0.29$. Ground truth score is obtained using original records from the dataset.}
  \label{tab:a2amos}
  \centering
  \begin{tabular}{l|ccccc|ccccc}
    \toprule
\multirow{2}{*}{\textbf{Model}} & \multicolumn{5}{|c|}{\textbf{Any-to-Any Voice Quality $\uparrow$}} & \multicolumn{5}{c}{\textbf{Any-to-Any Similarity $\uparrow$}} \\
& \textbf{F2F} & \textbf{F2M} & \textbf{M2M} & \textbf{M2F} & \textbf{Mean} & \textbf{F2F} & \textbf{F2M} & \textbf{M2M} & \textbf{M2F} & \textbf{Mean} \\
\midrule
Ground Truth & 4.27 & N/A & 4.47 & N/A & 4.37 & 4.39 & N/A & 4.17 & N/A & 4.28 \\
\midrule
AutoVC\cite{qian2019autovc} & 2.08 & 1.61 & 1.64 & 2.03 & 1.84 & 1.59 & 1.66 & 1.73 & 1.47 & 1.61 \\
VQMIVC\cite{wang2021vqmivc} & 3.64 & 3.73 & 3.67 & 3.70 & 3.69 & 1.96 & 2.23 & 2.22 & 1.95 & 2.09 \\
NVC-Net\cite{nguyen2021nvcnet} & 3.68 & 3.41 & 3.64 & 3.42 & 3.54 & 2.22 & 1.82 & 1.82 & 2.06 & 1.98 \\
\midrule
HiFi-VC (Proposed) & \bf 4.00 & \bf 3.98 & \bf 4.06 & \bf 4.09 & \bf 4.03 & \bf 3.50 & \bf 2.54 & \bf 2.70 & \bf 3.34 & \bf 3.02 \\

\bottomrule
\end{tabular}
\end{table*}

\begin{table}[ht]
  \caption{Any-to-any word error rate (WER), character error rate (CER) and prosody quality in terms of Pearson correlation coefficient (PCC) between source and predicted pitch features.}
  \label{tab:a2aobjective}
  \centering
  \begin{tabular}{l|ccc}
    \toprule
\multirow{2}{*}{\textbf{Model}} & \multicolumn{3}{|c}{\textbf{Any-to-Any}} \\
& \textbf{WER (\%) $\downarrow$} & \textbf{CER (\%) $\downarrow$} & \textbf{PCC $\uparrow$} \\
\midrule
AutoVC\cite{qian2019autovc} & 95.4 & 67.6 & 0.12 \\
VQMIVC\cite{wang2021vqmivc} &  32.2 & 16.7 & 0.55 \\
NVC-Net\cite{nguyen2021nvcnet} &  32.5 & 16.5 & 0.12 \\
\midrule
HiFi-VC (Proposed) & \bf 10.3 & \bf 4.2 & \bf 0.66 \\

\bottomrule
\end{tabular}
\end{table}

\section{Proposed Method}
Our model is based on encoder-decoder architecture with an additional speaker embedder, as shown in Figure \ref{fig:inference}. The proposed approach is inspired by TTS Skins \cite{polyak2019ttsskins}, NVC-Net \cite{nguyen2021nvcnet} and BNE-Seq2seqMoL \cite{liu2021f0prediction}.

\subsection{Model Architecture}

{\bf Linguistic encoder.} The goal of this encoder is to extract speaker-agnostic content information from the source voice sample. Our encoder is based on the TTS Skins approach \cite{polyak2019ttsskins}, as we make use of bottleneck features from an automated speech recognition (ASR) model. In our linguistic encoder, we apply the Conformer \cite{gulati2020conformer} ASR model pretrained by NVIDIA and available on the official website \footnote{\url{https://catalog.ngc.nvidia.com/orgs/nvidia/teams/nemo/models/stt_en_conformer_ctc_large_ls}}. The usage of the pretrained ASR allows us to largely speed up training and improve generalization as ASR representations are independent of the particular VC dataset.

{\bf F0 encoder.} As ASR is trained to extract linguistic information, it is not very accurate at capturing prosody. To overcome this limitation, we add a fundamental frequency (F0) predictor similar to BNE-Seq2seqMoL any-to-many VC approach \cite{liu2021f0prediction}. We also extract the boolean vocalization feature, which indicates regions where F0 cannot be estimated. Our F0 predictor consists of the WORLD extractor \cite{morise2016world} and a fully-convolutional network with 3 layers and instance normalization \cite{ulyanov2016instancenorm}, similar to BNE-Seq2seqMoL \cite{liu2021f0prediction}. The goal of instance normalization is to exclude speaker information from F0, while trainable subnetwork provides more flexibility in F0 coding.

{\bf Speaker embedder}. Any-to-any voice conversion requires using a speaker embedder network. Speaker embedder predicts the distribution of speaker feature vectors from an audio sample using a 5-layer residual fully-connected network. The multivariate normal distribution is defined by the mean vector and diagonal covariance matrix. During training actual speaker embedding is obtained via sampling, while in testing mean is used. Our architecture is similar to that used in NVC-Net \cite{nguyen2021nvcnet}. The speaker embedder is trained along with other modules in a single pipeline.

{\bf Decoder}. Our novel decoding approach extends previous GAN-based methods. Many previous works utilize HiFi GAN \cite{kong2020hifi} as a vocoder for the final decoding step. The goal of a vocoder is to convert a Mel-spectrogram into a waveform. While previous HiFi GAN-based approaches use different networks for decoder and vocoder, we combine both into a single module. We implement conditional HiFi GAN, with conditions obtained from the speaker embedder. ASR bottleneck and F0 features are directly served as GAN input.
In general, our approach simplifies predictor and provides better conditioning on the speaker embedding.

\subsection{Training Objectives}
During training, we freeze the ASR model and simultaneously optimize parameters of the F0 encoder, speaker embedder, decoder network and GAN discriminators. We do this by combining modified HiFi GAN losses \cite{kong2020hifi} with speaker embedder regularization loss from NVC-Net \cite{nguyen2021nvcnet}.

In our method, we do not use intermediate Mel-spectrogram representation. In contrast to the original HiFi GAN approach, we compute reconstruction loss between spectrograms of the source and predicted voice samples. Suppose $M_{source}$ is a Mel-spectrogram of the source and $M_{pred}$ is a spectrogram of the predicted waveform. Here, reconstruction loss would be defined as
\begin{equation}
    \mathcal{L}_{Rec} = ||M_{source} - M_{pred}||_1.
\end{equation}

HiFi GAN uses adversarial loss to make the predicted waveform indistinguishable from the natural one. Suppose $s$ is the predicted waveform, $x$ is the natural one and $D$ is a discriminator network. Then, the loss used for predictor optimization is defined as
\begin{equation}
    \mathcal{L}_{AdvP}(s) = (D(s) - 1)^2,
\end{equation}
and the loss for discriminator optimization is defined as
\begin{equation}
    \mathcal{L}_{AdvD}(s, x) = (D(x) - 1)^2 + (D(s))^2.
\end{equation}
GAN training is stabilized using feature matching loss. Suppose an output of the $i$-th discriminator layer is an $N_i$-dimensional vector $D_i$. In this case, feature matching loss is computed as
\begin{equation}
    \mathcal{L}_{FM}(s, x) = \sum\limits_{i=1}^{T}\frac{1}{N_i}||D_i(x) - D_i(s)||_1.
\end{equation}

In addition to the above, there is also speaker embedding regularization loss, similar to that in NVC-Net\cite{nguyen2021nvcnet}. Suppose, there is a reference record $r$. If we define mean and diagonal covariance matrix predicted by speaker embedder as $\mu(r)$ and $\Sigma(r)$, then regularization loss would be defined as
\begin{equation}
    \mathcal{L}_{Spk}(r) = \mathbb{D}_{KL}(\mathcal{N}(x;\mu(r),\Sigma(r))|| \mathcal{N}(x;0,I)).
\end{equation}

The final loss for predictor is a weighted sum of the losses described above:
\begin{multline}
    \mathcal{L}_{P} = \lambda_{Rec}\mathcal{L}_{Rec} + \lambda_{AdvP}\mathcal{L}_{AdvP} + \\ \lambda_{FM}\mathcal{L}_{FM} + \lambda_{Spk}\mathcal{L}_{Spk}.
\end{multline}
We use $\lambda_{Rec} = 45$, $\lambda_{AdvP} = 1$, $\lambda_{FM} = 1$, $\lambda_{Spk} = 0.01$.


\subsection{Implementation Details}
During training, all voice samples are converted to 24kHz. ASR produces features with a 40ms period and the F0 predictor generates features every 10ms. The F0 prediction network is designed to downsample F0 features to match those of ASR.

We slightly modify HiFi GAN to match the ASR feature frequency. We do this by increasing upsampling level from 256 to 960. We also remove a bias parameter from the final convolutional layer to stabilize mixed-precision training \cite{kim2021conditional}. We train our network for 120 epochs using Adam optimizer. We set initial learning rate to 0.0002 and use exponential scheduler with $\gamma = 0.995$. Each epoch takes 85 minutes on a single NVIDIA V100 GPU.


\section{Experiments}
In this section, we describe the experimental setup and our results, produced by subjective and objective evaluations. Voice conversion examples of the proposed method are provided in supplementary materials.

\subsection{Experimental Setup}
We use the VCTK dataset \cite{liu2019vctk} for training baselines and our model. The dataset includes 44242 voice samples from 110 speakers, among which are 47 male, 61 female, and 2 speakers of unknown gender.
We keep 6 speakers for any-to-any evaluation and use the other speakers during training. Voice samples do not overlap between training and testing.

We compare our method to baselines with publicly available implementations. The set of baselines we use includes AutoVC \cite{qian2019autovc}, VQMIVC \cite{wang2021vqmivc} and NVC-Net \cite{nguyen2021nvcnet}. The original AutoVC implementation failed to train in our setup, which is why we used an official pretrained AutoVC model for all comparisons. We applied the same AutoVC model to many-to-many and any-to-any tasks.

Voice conversion evaluation usually involves estimation of the output voice quality, voice similarity, linguistic and prosody consistency. We evaluate voice quality using subjective mean opinion score (MOS) \cite{ribeiro2011crowdmos}. Each predicted record is estimated with a grade ranging from 1 (completely unnatural) to 5 (completely natural).
We use a crowd-sourcing platform for markup. During markup, we place 11 entries on each page of tasks, including one honeypot, two examples for each possible combination of source and target speaker genders (M2M, F2M, M2F, F2F), as well as two original female and male voice samples. Each algorithm was evaluated by 400 samples in many-to-many setup and by 144 samples in any-to-any. Grades were obtained from 900 assessors with overlap 10 for many-to-many and 20 for any-to-any.

For voice similarity evaluation, we construct pairs of voice samples. Assessors are required to measure voice similarity between samples with grades between 1 (different) and 5 (same). We report the final MOS for different source and reference gender combinations.

Linguistic consistency is measured as word error rate (WER) and character error rate (CER) between source and output speech samples. As our method involves ASR features, WER and CER can be biased. To solve this problem, we use a different ASR model for evaluation\cite{silero}. We also report prosody consistency computed via Pearson correlation coefficient (PCC) between F0 tracks from source and predicted samples.

\subsection{Conversion Quality}
We perform separate evaluations for many-to-many and any-to-any setups. 
Voice quality and similarity MOS metrics for these two tasks are reported in \mbox{Table \ref{tab:mos}} and Table \ref{tab:a2amos} respectively. Among baselines, VQMIVC generally performs better in terms of voice quality, and NVC-Net achieves higher similarity. At the same time, the proposed HiFi-VC method outperforms all considered baselines both in terms of voice quality and similarity.

We also perform a set of objective evaluations aimed at linguistic and prosody consistency. Evaluation results for many-to-many and any-to-any tasks are reported in Table \ref{tab:objective} and Table \ref{tab:a2aobjective} respectively. Among baselines, NVC-Net achieves lower WER and CER in the many-to-many task while being on par with VQMIVC in the any-to-any task. On the other hand, VQMIVC achieves higher pitch consistency compared to other baselines. The proposed HiFi-VC method achieves the lowest WER and CER among all methods. At the same time, prosody predicted by HiFi-VC better correlates with the source speech sample.


\subsection{Discussion}
Experimental results suggest that HiFi GAN can be used either as a vocoder or as a joint decoder-vocoder module. The proposed HiFi-VC method achieves better conversion results than the baselines in both many-to-many and any-to-any setups. The proposed speaker conditioning scheme increases speaker similarity and the usage of ASR bottleneck and F0 features leads to high linguistic and prosody consistency.

\section{Conclusion}
In this work, we presented a novel voice conversion system, which combines the ASR model with direct waveform prediction using a conditioned HiFi GAN. A robust ASR feature extractor along with a speaker embedder allows this method to solve general any-to-any conversion tasks. According to multiple experiments involving subjective and objective evaluation, our method achieves better voice conversion than the baselines in terms of voice quality, similarity and consistency.

\clearpage 

\bibliographystyle{IEEEtran}

\bibliography{vc}

\begin{thebibliography}{10}
\providecommand{\url}[1]{#1}
\csname url@samestyle\endcsname
\providecommand{\newblock}{\relax}
\providecommand{\bibinfo}[2]{#2}
\providecommand{\BIBentrySTDinterwordspacing}{\spaceskip=0pt\relax}
\providecommand{\BIBentryALTinterwordstretchfactor}{4}
\providecommand{\BIBentryALTinterwordspacing}{\spaceskip=\fontdimen2\font plus
\BIBentryALTinterwordstretchfactor\fontdimen3\font minus
  \fontdimen4\font\relax}
\providecommand{\BIBforeignlanguage}[2]{{%
\expandafter\ifx\csname l@#1\endcsname\relax
\typeout{** WARNING: IEEEtran.bst: No hyphenation pattern has been}%
\typeout{** loaded for the language `#1'. Using the pattern for}%
\typeout{** the default language instead.}%
\else
\language=\csname l@#1\endcsname
\fi
#2}}
\providecommand{\BIBdecl}{\relax}
\BIBdecl

\bibitem{ning2019ttsreview}
Y.~Ning, S.~He, Z.~Wu, C.~Xing, and L.-J. Zhang, ``A review of deep learning
  based speech synthesis,'' \emph{Applied Sciences}, vol.~9, no.~19, p. 4050,
  2019.

\bibitem{sisman2020overview}
B.~Sisman, J.~Yamagishi, S.~King, and H.~Li, ``An overview of voice conversion
  and its challenges: From statistical modeling to deep learning,''
  \emph{IEEE/ACM Transactions on Audio, Speech, and Language Processing},
  vol.~29, pp. 132--157, 2020.

\bibitem{luo2020singing}
Y.-J. Luo, C.-C. Hsu, K.~Agres, and D.~Herremans, ``Singing voice conversion
  with disentangled representations of singer and vocal technique using
  variational autoencoders,'' in \emph{ICASSP 2020-2020 IEEE International
  Conference on Acoustics, Speech and Signal Processing (ICASSP)}.\hskip 1em
  plus 0.5em minus 0.4em\relax IEEE, 2020, pp. 3277--3281.

\bibitem{nakamura2012speakingaid}
K.~Nakamura, T.~Toda, H.~Saruwatari, and K.~Shikano, ``Speaking-aid systems
  using gmm-based voice conversion for electrolaryngeal speech,'' \emph{Speech
  communication}, vol.~54, no.~1, pp. 134--146, 2012.

\bibitem{shahnawazuddin2020vcaugmentation}
S.~Shahnawazuddin, N.~Adiga, K.~Kumar, A.~Poddar, and W.~Ahmad, ``Voice
  conversion based data augmentation to improve children's speech recognition
  in limited data scenario.'' in \emph{Interspeech}, 2020, pp. 4382--4386.

\bibitem{maouche2020anonymization}
M.~Maouche, B.~M.~L. Srivastava, N.~Vauquier, A.~Bellet, M.~Tommasi, and
  E.~Vincent, ``A comparative study of speech anonymization metrics,'' in
  \emph{INTERSPEECH 2020}, 2020.

\bibitem{qian2019autovc}
K.~Qian, Y.~Zhang, S.~Chang, X.~Yang, and M.~Hasegawa-Johnson, ``Autovc:
  Zero-shot voice style transfer with only autoencoder loss,'' in
  \emph{International Conference on Machine Learning}.\hskip 1em plus 0.5em
  minus 0.4em\relax PMLR, 2019, pp. 5210--5219.

\bibitem{wang2021vqmivc}
D.~Wang, L.~Deng, Y.~T. Yeung, X.~Chen, X.~Liu, and H.~Meng, ``Vqmivc: Vector
  quantization and mutual information-based unsupervised speech representation
  disentanglement for one-shot voice conversion,'' \emph{arXiv preprint
  arXiv:2106.10132}, 2021.

\bibitem{peddinti2015tdnn}
V.~Peddinti, D.~Povey, and S.~Khudanpur, ``A time delay neural network
  architecture for efficient modeling of long temporal contexts,'' in
  \emph{Sixteenth annual conference of the international speech communication
  association}, 2015.

\bibitem{ravanelli2018speaker}
M.~Ravanelli and Y.~Bengio, ``Speaker recognition from raw waveform with
  sincnet,'' in \emph{2018 IEEE Spoken Language Technology Workshop
  (SLT)}.\hskip 1em plus 0.5em minus 0.4em\relax IEEE, 2018, pp. 1021--1028.

\bibitem{schneider2019wav2vec}
S.~Schneider, A.~Baevski, R.~Collobert, and M.~Auli, ``wav2vec: Unsupervised
  pre-training for speech recognition,'' \emph{arXiv preprint
  arXiv:1904.05862}, 2019.

\bibitem{oord2016wavenet}
A.~v.~d. Oord, S.~Dieleman, H.~Zen, K.~Simonyan, O.~Vinyals, A.~Graves,
  N.~Kalchbrenner, A.~Senior, and K.~Kavukcuoglu, ``Wavenet: A generative model
  for raw audio,'' \emph{arXiv preprint arXiv:1609.03499}, 2016.

\bibitem{polyak2019ttsskins}
A.~Polyak, L.~Wolf, and Y.~Taigman, ``Tts skins: Speaker conversion via asr,''
  \emph{arXiv preprint arXiv:1904.08983}, 2019.

\bibitem{mottini2021voicy}
A.~Mottini, J.~Lorenzo-Trueba, S.~V.~K. Karlapati, and T.~Drugman, ``Voicy:
  Zero-shot non-parallel voice conversion in noisy reverberant environments,''
  \emph{arXiv preprint arXiv:2106.08873}, 2021.

\bibitem{nguyen2021nvcnet}
B.~Nguyen and F.~Cardinaux, ``Nvc-net: End-to-end adversarial voice
  conversion,'' \emph{arXiv preprint arXiv:2106.00992}, 2021.

\bibitem{li2021starganv2}
Y.~A. Li, A.~Zare, and N.~Mesgarani, ``Starganv2-vc: A diverse, unsupervised,
  non-parallel framework for natural-sounding voice conversion,'' \emph{arXiv
  preprint arXiv:2107.10394}, 2021.

\bibitem{prenger2019waveglow}
R.~Prenger, R.~Valle, and B.~Catanzaro, ``Waveglow: A flow-based generative
  network for speech synthesis,'' in \emph{ICASSP 2019-2019 IEEE International
  Conference on Acoustics, Speech and Signal Processing (ICASSP)}.\hskip 1em
  plus 0.5em minus 0.4em\relax IEEE, 2019, pp. 3617--3621.

\bibitem{kong2020hifi}
J.~Kong, J.~Kim, and J.~Bae, ``Hifi-gan: Generative adversarial networks for
  efficient and high fidelity speech synthesis,'' \emph{Advances in Neural
  Information Processing Systems}, vol.~33, pp. 17\,022--17\,033, 2020.

\bibitem{kawahara1999straight}
H.~Kawahara, I.~Masuda-Katsuse, and A.~De~Cheveigne, ``Restructuring speech
  representations using a pitch-adaptive time--frequency smoothing and an
  instantaneous-frequency-based f0 extraction: Possible role of a repetitive
  structure in sounds,'' \emph{Speech communication}, vol.~27, no. 3-4, pp.
  187--207, 1999.

\bibitem{perraudin2013fastgriffin}
N.~Perraudin, P.~Balazs, and P.~L. S{\o}ndergaard, ``A fast griffin-lim
  algorithm,'' in \emph{2013 IEEE Workshop on Applications of Signal Processing
  to Audio and Acoustics}.\hskip 1em plus 0.5em minus 0.4em\relax IEEE, 2013,
  pp. 1--4.

\bibitem{sun2016manyone}
L.~Sun, K.~Li, H.~Wang, S.~Kang, and H.~Meng, ``Phonetic posteriorgrams for
  many-to-one voice conversion without parallel data training,'' in \emph{2016
  IEEE International Conference on Multimedia and Expo (ICME)}.\hskip 1em plus
  0.5em minus 0.4em\relax IEEE, 2016, pp. 1--6.

\bibitem{liu2021vqbn}
Y.~Liu, C.~Yu, W.~Shuai, Z.~Yang, Y.~Chao, and W.~Zhang, ``Non-parallel
  any-to-many voice conversion by replacing speaker statistics,'' \emph{Proc.
  Interspeech 2021}, pp. 1369--1373, 2021.

\bibitem{kim2021assem}
K.-w. Kim, S.-w. Park, J.~Lee, and M.-c. Joe, ``Assem-vc: Realistic voice
  conversion by assembling modern speech synthesis techniques,'' \emph{arXiv
  preprint arXiv:2104.00931}, 2021.

\bibitem{chou2019adain}
J.-c. Chou, C.-c. Yeh, and H.-y. Lee, ``One-shot voice conversion by separating
  speaker and content representations with instance normalization,''
  \emph{arXiv preprint arXiv:1904.05742}, 2019.

\bibitem{sakamoto2021stargan}
S.~Sakamoto, A.~Taniguchi, T.~Taniguchi, and H.~Kameoka, ``Stargan-vc+ asr:
  Stargan-based non-parallel voice conversion regularized by automatic speech
  recognition,'' \emph{arXiv preprint arXiv:2108.04395}, 2021.

\bibitem{chen2021tvqvc}
Z.~Chen and P.~Zhang, ``Tvqvc: Transformer based vector quantized variational
  autoencoder with ctc loss for voice conversion,'' \emph{Proc. Interspeech
  2021}, pp. 826--830, 2021.

\bibitem{liu2021f0prediction}
S.~Liu, Y.~Cao, D.~Wang, X.~Wu, X.~Liu, and H.~Meng, ``Any-to-many voice
  conversion with location-relative sequence-to-sequence modeling,''
  \emph{IEEE/ACM Transactions on Audio, Speech, and Language Processing},
  vol.~29, pp. 1717--1728, 2021.

\bibitem{gulati2020conformer}
A.~Gulati, J.~Qin, C.-C. Chiu, N.~Parmar, Y.~Zhang, J.~Yu, W.~Han, S.~Wang,
  Z.~Zhang, Y.~Wu \emph{et~al.}, ``Conformer: Convolution-augmented transformer
  for speech recognition,'' \emph{arXiv preprint arXiv:2005.08100}, 2020.

\bibitem{morise2016world}
M.~Morise, F.~Yokomori, and K.~Ozawa, ``World: a vocoder-based high-quality
  speech synthesis system for real-time applications,'' \emph{IEICE
  TRANSACTIONS on Information and Systems}, vol.~99, no.~7, pp. 1877--1884,
  2016.

\bibitem{ulyanov2016instancenorm}
D.~Ulyanov, A.~Vedaldi, and V.~Lempitsky, ``Instance normalization: The missing
  ingredient for fast stylization,'' \emph{arXiv preprint arXiv:1607.08022},
  2016.

\bibitem{kim2021conditional}
J.~Kim, J.~Kong, and J.~Son, ``Conditional variational autoencoder with
  adversarial learning for end-to-end text-to-speech,'' in \emph{International
  Conference on Machine Learning}.\hskip 1em plus 0.5em minus 0.4em\relax PMLR,
  2021, pp. 5530--5540.

\bibitem{liu2019vctk}
Z.~Liu and B.~Mak, ``Cross-lingual multi-speaker text-to-speech synthesis for
  voice cloning without using parallel corpus for unseen speakers,''
  \emph{arXiv preprint arXiv:1911.11601}, 2019.

\bibitem{ribeiro2011crowdmos}
F.~Ribeiro, D.~Flor{\^e}ncio, C.~Zhang, and M.~Seltzer, ``Crowdmos: An approach
  for crowdsourcing mean opinion score studies,'' in \emph{2011 IEEE
  international conference on acoustics, speech and signal processing
  (ICASSP)}.\hskip 1em plus 0.5em minus 0.4em\relax IEEE, 2011, pp. 2416--2419.

\bibitem{silero}
S.~Team, ``Silero models: pre-trained enterprise-grade stt / tts models and
  benchmarks,'' \url{https://github.com/snakers4/silero-models}, 2022.

\end{thebibliography}

\end{document}